\documentclass[5p]{elsarticle}

\usepackage{dblfloatfix} 
\usepackage{float}
\usepackage{graphicx}

\usepackage{caption}

\journal{arXiv.org Preprint}

\begin{document}

\begin{frontmatter}

\title{Investigation of geometrically necessary dislocation structures in compressed Cu micropillars by 3-dimensional HR-EBSD}

\author[mymainaddress,mysecondaryaddress]{Szilvia Kal\'{a}cska\corref{mycorrespondingauthor}}
\cortext[mycorrespondingauthor]{Corresponding author}
\ead{szilvia.kalacska@empa.ch}
\author[mymainaddress]{Zolt\'{a}n Dankh\'{a}zi}
\author[mymainaddress]{Gyula Zilahi}
\author[mysecondaryaddress]{Xavier Maeder}
\author[mysecondaryaddress]{Johann Michler}
\author[mymainaddress]{P\'{e}ter Dus\'{a}n Isp\'{a}novity}
\author[mymainaddress]{Istv\'{a}n Groma}

\address[mymainaddress]{E\"{o}tv\"{o}s Lor\'{a}nd University, Department of Materials Physics, 1117 Budapest, P\'{a}zm\'{a}ny P\'{e}ter s\'{e}tany 1/a. Hungary}
\address[mysecondaryaddress]{Empa, Swiss Federal Laboratories for Materials Science and Technology, Laboratory of Mechanics of Materials and Nanostructures, CH-3602 Thun, Feuerwerkerstrasse 39. Switzerland}

\begin{abstract}
Mechanical testing of micropillars is a field that involves new physics, as the behaviour of materials is non-deterministic at this scale. To better understand their deformation mechanisms we applied 3-dimensional high angular resolution electron backscatter diffraction (3D HR-EBSD) to reveal the dislocation distribution in deformed single crystal copper micropillars. Identical micropillars ($6$ $\mu$m $\times 6$ $\mu$m $\times 18$ $\mu$m in size) were fabricated by focused ion beam (FIB) and compressed at room temperature. The deformation process was stopped at different strain levels ($\approx$ $1\%$, $4\%$ and $10\%$) to study the evolution of geometrically necessary dislocations (GNDs). Serial slicing with FIB and consecutive HR-EBSD mapping on the (100) side was used to create and compare 3-dimensional maps of the deformed volumes. Average GND densities were calculated for each deformation step. Total dislocation density calculation based on X-ray synchrotron measurements were conducted on the $4\%$ pillar to compare dislocation densities determined by the two complementary methods. Scanning transmission electron microscopy (STEM) and transmission electron microscopy (TEM) images were captured on the $10\%$ pillar to visualize the actual dislocation structure. With the 3D HR-EBSD technique we have studied the geometrically necessary dislocations evolving during the deformation of micropillars. An intermediate behaviour was found at the studied sample size between bulk and nanoscale plasticity: A well-developed dislocation cell structure built up upon deformation but with significantly lower GND density than in bulk. This explains the simultaneous observation of strain hardening and size effect at this scale.

\end{abstract}


\end{frontmatter}


\section{Introduction}

Understanding the mechanisms during the plastic deformation of crystalline materials is a central problem in materials science. In the last fifteen years one of the remarkable findings was that plastic deformation of crystals becomes dramatically different when the sample size is reduced to the micron or submicron scale, compared to the behaviour of bulk materials \cite{Uchic2004}. This difference decisively influences today’s industrial sectors with focus on miniaturization, as the size reduction effect is no longer negligible. Mechanical testing of micropillars require new physics to describe the material’s unique response to deformation. Traditional deterministic approaches of plasticity cannot be applied as the stress-strain behaviour varies from sample to sample.

Microstructure formed by plastic deformation can not only be influenced by the mode of deformation (for example compression, tension, or torsion has its own texture typical of the process) but by the size of the sample, too. During micropillar compression dislocation nucleation, dislocation-dislocation interaction and their collective avalanche-like motion control the deformation process \cite{Papanikolaou2018}. Although many attempts have been made to characterize micromechanical behaviour, complete comprehension of mechanical testing of micron-sized pillars are still missing due to the fact that we have limited possibilities to perform 3D measurements at such small scales. The aim of this paper is to investigate the dislocation microstructure developing in the sample that enables non-deterministic stress-strain behaviour to occur.

Micromechanical tests on micropillars were widely performed to monitor size-dependent stress-strain behaviour \cite{Frick2008, Dehm2018} and to study slip system activation and dislocation distribution \cite{Norfleet2008, Schneider2013}. In order to access information from the interior of the deformed structures, experimental results were complemented by discrete dislocation dynamic simulations \cite{Csikor2007, Kondori2017}.

Kiener \emph{et al.} \cite{Kiener2011} have suggested from experiments and simulations that the size effect observed in work hardening of micropillars originates from the build-up of geometrically necessary dislocations (GNDs) in the pillars. It has been recently shown by STEM measurements conducted on copper single crystal samples \cite{Zhao2019} that there exists a critical size for micropillars where complex dislocation structure can appear. The study showed that pillars of 5 $\mu$m in diameter above $8\%$ compressive strain will form dislocation cell structures which have a characteristic length-scale of about $0.5-1$ $\mu$m. This characteristic value corresponds well to the dislocation cell size identified by high angular resolution electron backscatter diffraction (HR-EBSD) in bulk copper single crystals \cite{Kalacska2017}. Therefore, if micropillars are fabricated and investigated around this critical dimension, we can study the distribution of GNDs in a well-defined volume, where the size-dependent mechanical behaviour will start to differ from the bulk case.

The formation of the cell structure in Cu single crystals have been known for decades \cite{Staker1972, Prinz1980, Knoesen1982}, but its' presence in compressed micropillars is somewhat surprising since the cell size is in the order of magnitude of $\mu$m. It is natural to assume that if for very small sample the deformation mechanism is controlled by dislocation starvation \cite{Greer2005} while at large enough (or bulk) specimen dislocations tend to form cells separated by relatively narrow walls, then at an intermediate size there exists a transition between the two mechanisms \cite{Gao2010}, resulting in a detectable GND cell structure with much lower dislocation density than in bulk. This intermittent scale where dislocation cell-formation start to appear has not yet been understood. Size dependent strengthening effects have been reported experimentally \cite{Geers2006, Greer2011, Hug2015}, but the underlying physical mechanism driving these size effects is still debated \cite{Liu2016}. The fact that the distribution of GNDs is also size dependent has been studied by discrete dislocation dynamics simulations \cite{Guruprasad2008, Papanikolaou2017, Kondori2017} and dislocation-based crystal plasticity modelling \cite{Lin2016}, but experimentally has never been determined before in 3 dimensions. To provide quantitative inputs for validating these numerical simulations we need to investigate the types and spatial morphology of GNDs accumulating in the system in different sized micropillars. Therefore, to shed light on small scale plasticity at an intermediate sample volume we aim to map GND distribution evolution at various deformation levels. 

Other techniques like HR-SEM or digital image cor\-re\-la\-tion-based imaging have been applied to observe slip systems and dislocation activities \cite{Choi2015}. These characterization methods can only be applied on the surface, giving limited information of what is going on in the whole volume of the material.

With the HR-EBSD technique the density of GNDs can be calculated \cite{Arsenlis1999, Wilkinson2010}. Geometrically necessary dislocations appear in the crystal to accommodate lattice curvature. These GNDs can be detected through the strain-gradient fields, so they are accessible by HR-EBSD. The rest of the dislocations in the material (so-called statistically stored dislocations) are invisible to this technique \cite{ElDasher2003, Pantleon2008}. The $\alpha_{ij}$ local dislocation density tensor was introduced by Nye \cite{Nye1953}, and it can be written as

\begin{eqnarray}\label{eq:01}
\alpha_{ij} = \sum_t b_i^t l_j^t\rho^t,
\end{eqnarray}
where dislocations are characterized by the Burgers vector $\mathbf{b}^t$ and  their line direction $\mathbf{l}^t$ for different $t$ types of dislocations. The sum is over all types of dislocations present in the material, and $\rho ^t$ denotes the dislocation density value of type $t$. Because backscattered electrons interact with the first ~20-50 nm of the surface \cite{Chen2011}, only the $\alpha_{i3}$ components of Nye's dislocation density tensor can be determined experimentally:

\begin{eqnarray}\label{eq:ai3}
\alpha_{i3} = \partial_1 \beta_{i2} - \partial_2 \beta_{i1}, i=1,2,3,
\end{eqnarray}
where $\beta_{ij}$ are the deformation gradient tensor components \cite{Wilkinson2009}. $\beta_{ij}$ values are calculated by HR-EBSD cross-correlation based evaluation. 

Using only the three established components $\alpha_{i3}$, reduced GND density values can be calculated as:

\begin{eqnarray}\label{eq:gnd_ai3}
\rho_{GND} = \frac{1}{|b|N}\sum_{N} \sqrt{\alpha_{13}^2 + \alpha_{23}^2 + \alpha_{33}^2},
\end{eqnarray}
where $N$ denotes the total number of points in the EBSD map. Studies have shown that EBSD-based GND density calculation is sensitive to the applied step size of the measurement \cite{Wright2015, Jiang2013}, that could be a limiting factor for the precision of $\rho_{GND}$. 

Another widely accepted procedure to give a lower bound estimate for the GND density can be described by utilizing an optimization method to minimize the total dislocation line energy ($L^1$ optimization scheme) \cite{Wilkinson2010}. Terms that cannot be measured are set to zero in this evaluation, and only pure edge or screw dislocations with the same magnitude of Burger’s vector are considered to be present in the crystal. These estimations lead to the possibility to distinguish edge and screw dislocations based on their energies:

\begin{eqnarray}\label{eq:02}
\frac{E_{edge}}{E_{screw}}=\frac{1}{1-\nu},
\end{eqnarray}
where $\nu$ is the Poisson number. 

In this article, we utilize both GND calculation techniques on the same dataset in order to calculate GND density distributions in copper single crystalline micropillars. We estimate the dislocation accumulation and the distribution evolution due to different compression levels. We aim to discuss differences in the applied calculation methods in detail. For this study, only the total GND density values were used, no distinction was made based on the types of dislocations in the system.

In order to learn about microstructural changes in a confined volume, FIB was utilized to fabricate samples with identical geometries \cite{Uchic2005}. FIB enables us to perform serial sectioning coupled with HR-EBSD measurements. After each FIB slice was performed, the newly cut surface was measured by HR-EBSD. All of the slices were then evaluated, and the slices were put together to form a 3D model of the mapped volume.

\section{Applied methods}

Orientation determination was done using conventional electron backscatter diffraction (EBSD). HR-EBSD uses cross-correlation evaluation on the diffraction patterns to calculate local strain and stress tensor components \cite{Britton2012}. The HR-EBSD technique requires a reference diffraction pattern for the comparison, which is ideally recorded in the strain-free state of the lattice. A perfect strain-free reference pattern can be experimentally hard to obtain, therefore either simulated patterns can be generated for this purpose \cite{Winkelman2007, Winkelman2012, Fullwood2015}, or a pattern with the presumably lowest stress is chosen, creating a relative scale for the results.

\subsection{Sample preparation and experimental realization}

\begin{figure}[!ht]
\begin{center}
\includegraphics[width=0.5\textwidth]{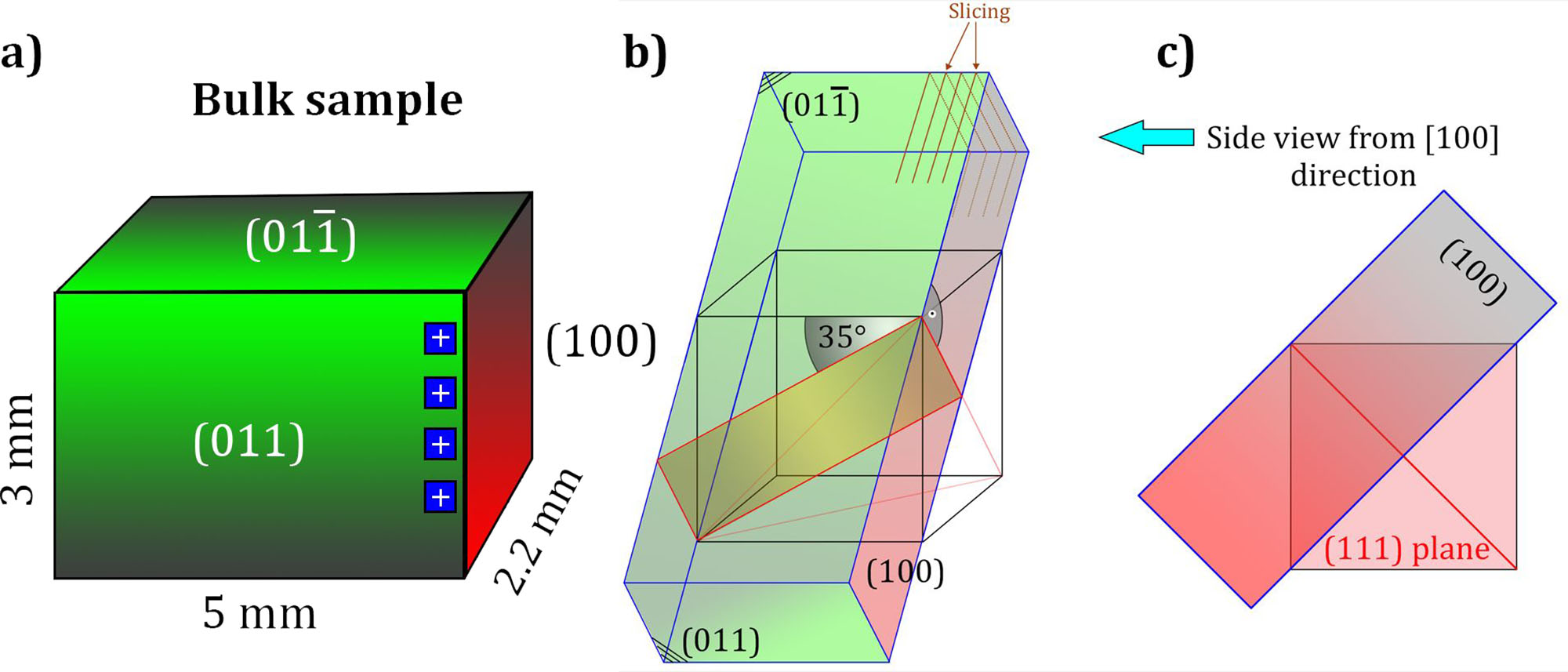} \\
\caption{\label{fig:01} a) Schematics of the bulk copper single crystal sample shown with its external dimensions. Colors are indicating the orientation of the surfaces. Blue squares represent the deposited Pt caps (not in proportion). b) Schematics of a single micropillar after fabrication. A possible \{111\} type slip plane is also indicated. c) The micropillar depicted from the EBSD point-of-view. (colours online)}
\end{center}
\end{figure}

For the experiment a previously heat-treated copper single crystal sample with a well-defined orientation was electropolished with Buehler’s D2 solution. The sample’s orientation and dimensions can be seen in Figure \ref{fig:01} a), where blue squares mark the position of the four initially created micropillars (viewed from the top). Each pillar had the same orientation depicted in Figure \ref{fig:01} b), where the later FIB slicing direction is also shown. The $[011]$ ideal double slip direction  was chosen as compression axis \cite{Raabe2007}.

\begin{figure}[!ht]
\begin{center}
\includegraphics[width=0.45\textwidth]{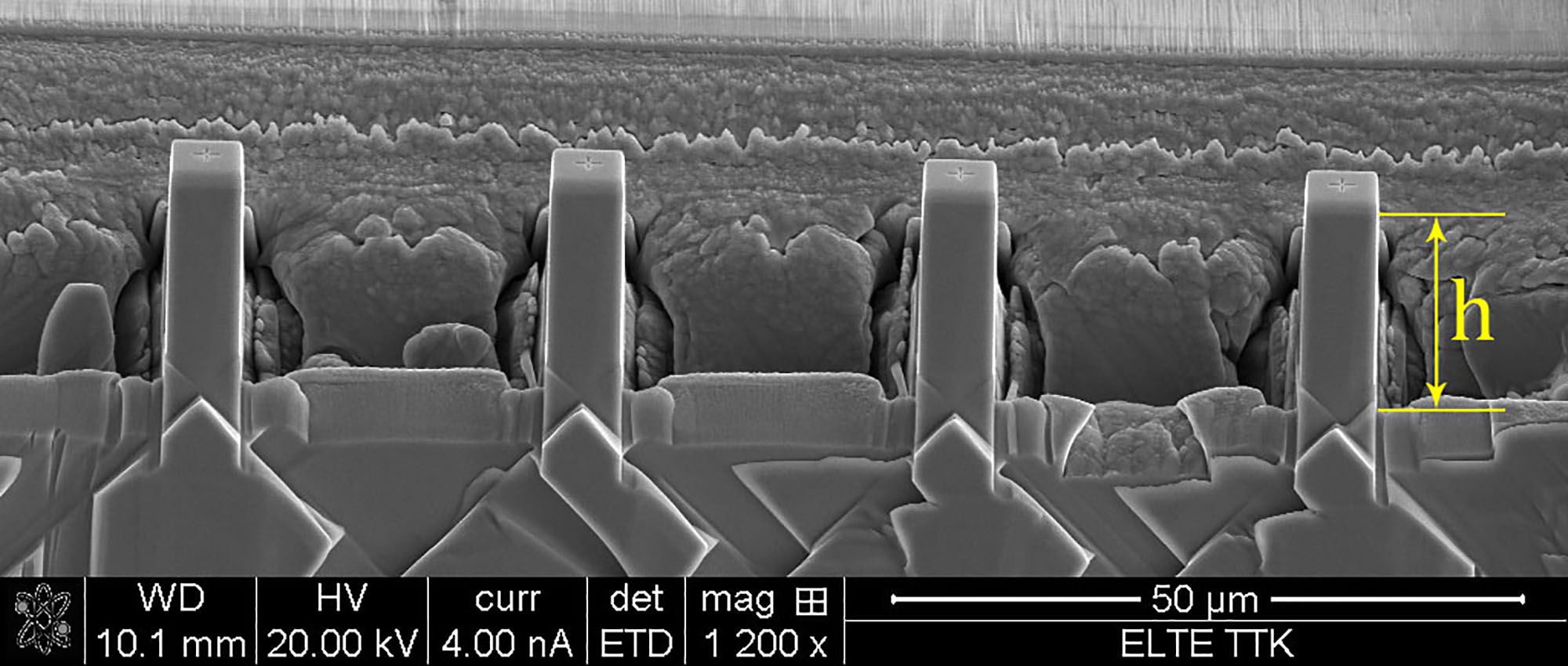} \\
\caption{\label{fig:02} The final four micropillars before compression. Pillar dimensions are approximately $6$ $\mu m \times 6$ $\mu m \times 18$ $\mu$m. $h$ indicates the height measurement borders.}
\end{center}
\end{figure}

The pillars were fabricated by a FEI Quanta 3D FIB-SEM system. As it is well known, FIB fabrication introduces surface defects due to Ga$^+$ ion bombardment \cite{Shim2009}. Rough milling was done in a lathe milling position \cite{Uchic2009}, using 30 kV ion beam with currents of 15 nA decreasing to 5 nA as the milling got closer to the protected volume. A final fabrication step with a beam of 30 kV and 0.5 nA was used to minimize the surface defects of the pillars and to improve EBSD pattern quality. Moreover, the top of the pillars were protected with nanocrystalline Pt caps \cite{Alaie2015} deposited by a gas injection system. The Pt caps helped positioning the nanoindenter's flat punch tip, and they also acted as a very hard buffer material between the pillar and the 10 $\mu$m wide flat punch tip during room temperature micromechanical testing. The resulting micropillars were $6$ $\mu$m $\times 6$ $\mu$m $\times 18$ $\mu$m in size. Their heights were measured from the edge of the FIB milled plateau until the Pt cap, as indicated in Figure \ref{fig:02} ($h$). The lathe milling setup enables pillar fabrication with very small taper angles. These pillars can be considered as non-tapered ($< 0.8 ^{\circ}$).

\begin{figure}[!ht]
\begin{center}
\includegraphics[width=0.45\textwidth]{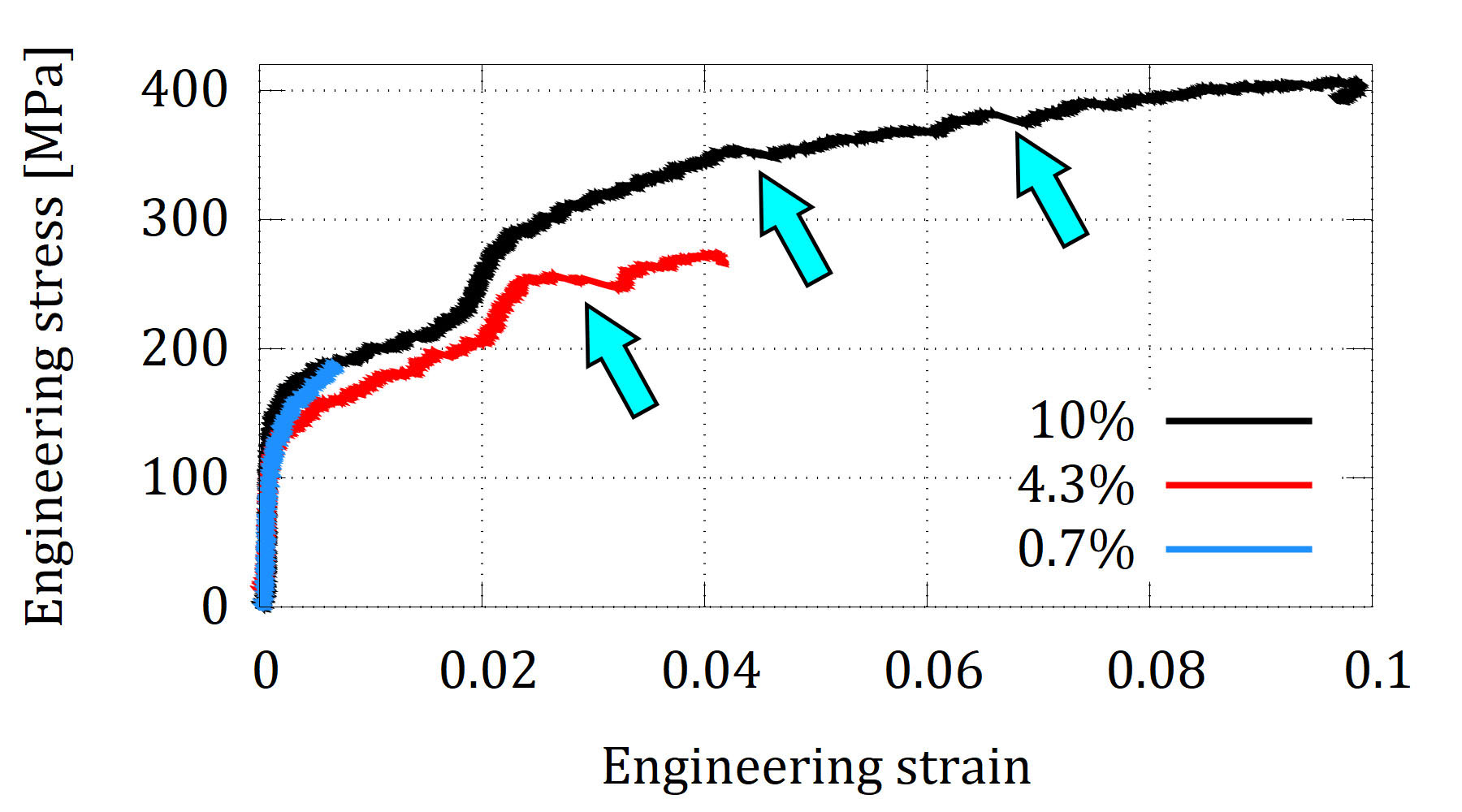} \\
\caption{\label{fig:03} Engineering stress plotted as a function of engineering strain for the three compressed micropillars deformed to three different strains. Blue arrows indicate load drops. (colours online)}
\end{center}
\end{figure}

After sample preparation, a custom-made nano\-in\-den\-ter \cite{Hegyi2017} was used. Three pillars were compressed to  0.7\%, 4.3\% and 10\% strains, respectively, with a fixed crosshead velocity of $9 \times 10^{-3}$ $\mu$m/s. The calculated engineering stress ($\sigma_E$) and engineering strain ($\varepsilon_E$) curves are plotted in Figure \ref{fig:03}. The $\sigma_E -\varepsilon_E$ curves contain a few load drops that are typical to uniaxial testing at the micron scale \cite{Dehm2018}, but their presence is usually more apparent at smaller pillar diameters. The magnitude of load drops depends on the size of the pillar, sample orientation (single or multiple slip), the speed of deformation (faster compression can induce bigger avalanche), etc.

As the sample was oriented for multiple slip, we can expect hardening at lower strain values \cite{Argon}.  EBSD orientation analysis on the (100) side of the pillars confirmed a small ($\sim 4^{\circ}$) misorientation from the exact direction, that is caused by the mounting of the bulk sample. Misalignment can also occur between the pillar axis and the compression direction due to the fact that between the fabrication and deformation process the sample was relocated. In case of compression, the above mentioned misorientations force the lattice to rotate during the initial loading stage, creating GNDs close to the flat punch tip – pillar interface \cite{Zhang2006}. Top coating can also eventuate higher hardening compared to uncoated samples at low strain values \cite{Kiener2011}.

Pillars deformed to $4.3\%$ and $10\%$ both show hardening at strain levels around $0.02$. This phenomenon could also be observed visually during compression, where a contrast change in the secondary electron image indicated the process (see in the supplementary video). The calculated reduced (with respect to the shear modulus) strain-hardening rates for the two pillars are in the typical order of $10^{-2}$ \cite{Argon}.

\begin{figure}[!ht]
\begin{center}
\includegraphics[width=0.4\textwidth]{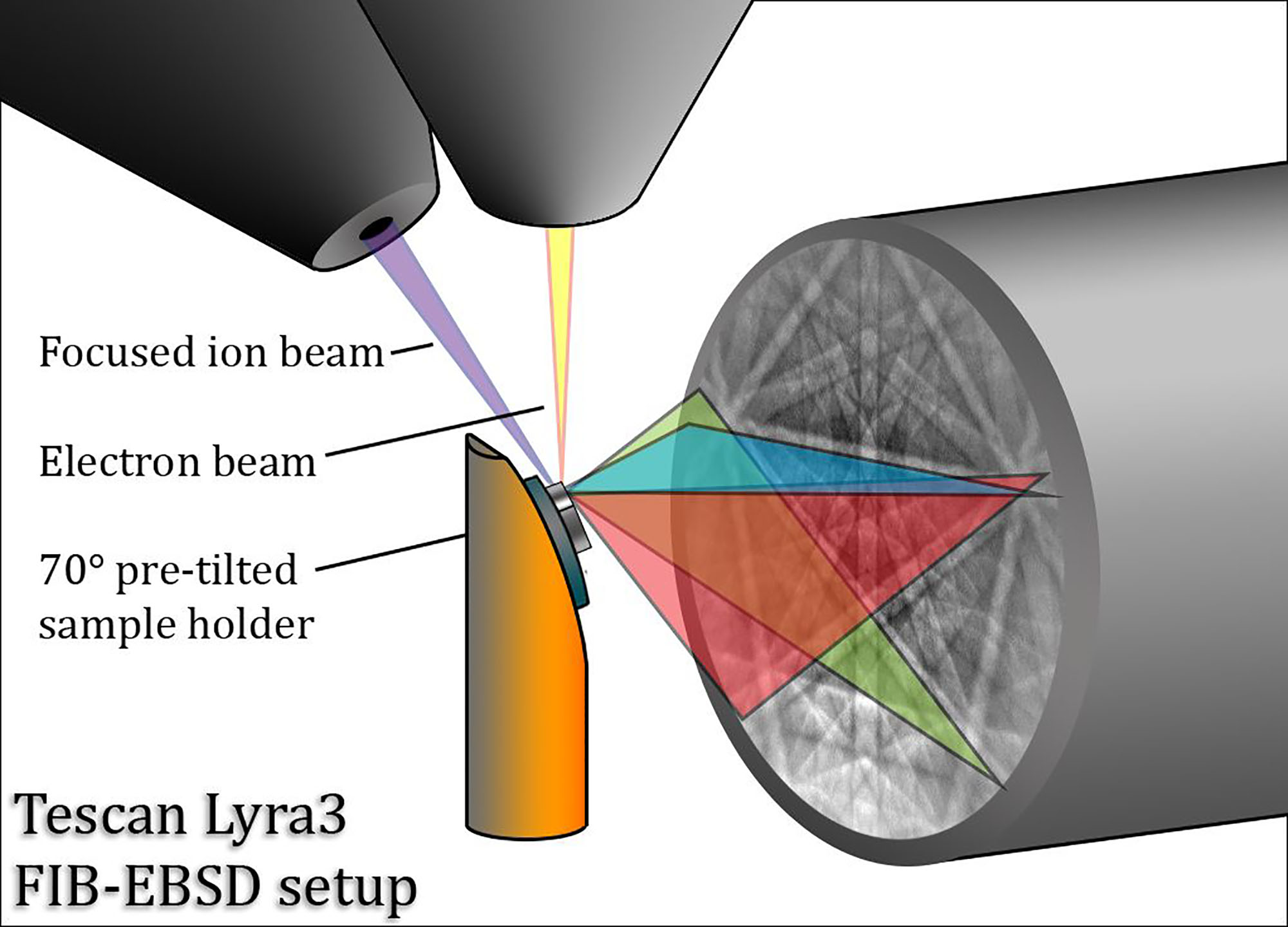} \\
\caption{\label{fig:04} Geometry in the Tescan Lyra3 SEM, where the chamber is specially designed to completely eliminate sample movement during 3D EBSD measurements. (colours online)}
\end{center}
\end{figure}

For the HR-EBSD measurements, two different systems were used (FEI Quanta 3D and Tescan Lyra3). In case of the FEI SEM, the EBSD camera and the FIB gun is mounted on the chamber in a way that sample rotation is always necessary between the FIB grazing incidence milling position \cite{West2009} and the EBSD mapping position. This setup is disadvantageous because after stage movement an additional time is needed to reduce stage drift (and hence inaccurate FIB slicing or drift during EBSD mapping). To completely remove this unwanted drift phenomena, the 0.7\% and 10\% deformed micropillars were investigated with a Tescan Lyra3 FIB-SEM system where the chamber is specially designed to allow rotation-free consecutive FIB slicing and EBSD mapping by an Edax DigiView camera. Figure \ref{fig:04} shows the geometry of the pillar slicing process. FIB slicing was done with an ion beam of 30 kV, $\sim 300$ pA. The conditions of the diffraction pattern collection are summarized in Table \ref{table:01}. EBSD measurements were recorded by OIM Data Collection v7, and analysed using OIM Analysis v7. For the cross-correlation based HR-EBSD analysis BLG Vantage CrossCourt v4 was used, that is based on Wilkinson’s evaluation method \cite{Wilkinson2006}. The $\alpha_{i3}$ components calculation was done by a C++ program developed by the Authors. Various number of slices were used to reconstruct the mapped volumes.

\begin{table*}[!ht] 
\begin{tabular}[width=\textwidth]{|l|l|l|l|l|l|}\hline 
System & EBSD beam & Binning & Working & Thickness & Mapped area\\
 & conditions &(image size) & distance & of one slice & (square grid) \\
\hline FEI Quanta 3D with & 20 kV, 4 nA  & $1\times 1$, & 15 mm & 135 nm  & 6.4 $\mu$m $\times$ 15 $\mu$m 
\\ 
Edax Hikari camera & (analytical  & (480 pixel $\times$  &  & ($\pm$15 nm) & 100 nm step
\\
(4.3\% deformation) & mode) &  480 pixel) &  & & 
\\

\hline Tescan Lyra3 with & 20 kV, 15 nA & $2\times 2$, & 9 mm & 90 nm & 7 $\mu$m $\times$ 16 $\mu$m
\\
Edax DigiView camera & & (442 pixel $\times$ & &($\pm$10 nm) & 100 nm step
\\
(0.7\%, 10\% deformation) & & 442 pixel) & & & 
\\
\hline

\end{tabular} 
\centering \caption{Beam conditions and mapping properties summarized for the two utilized SEM-FIB systems.} 
\label{table:01} 
\end{table*}

After the cross-correlation evaluation, calculated values of stress components, lattice rotations and GND density data were plotted using identical colour scaling. The magnitude of grayscale was then later interpreted as different values for the channel conversion and volume rendering in the 3D software. In the present work we applied Photoshop to manually align the slices by changing the visibility of one slice over the other. Amira 3D software was used to build up the 3D models from the 2D slices. 

In order to verify dislocation densities measured by HR-EBSD, one of the pillars were studied by X-ray diffraction (XRD). Line profile analysis is widely used to determine the total dislocation density value in bulk samples. By HR-EBSD, only a fraction of dislocations (GNDs) can be accessed, while XRD provides the total dislocation density value for the system. The widening of the measured profile is caused by dislocations. By linear fitting on the asymptotic region of the calculated moments the total dislocation density can be obtained. The method is explained in Ref. \cite{Groma2000}.

\begin{figure}[!ht]
\begin{center}
\includegraphics[width=0.45\textwidth]{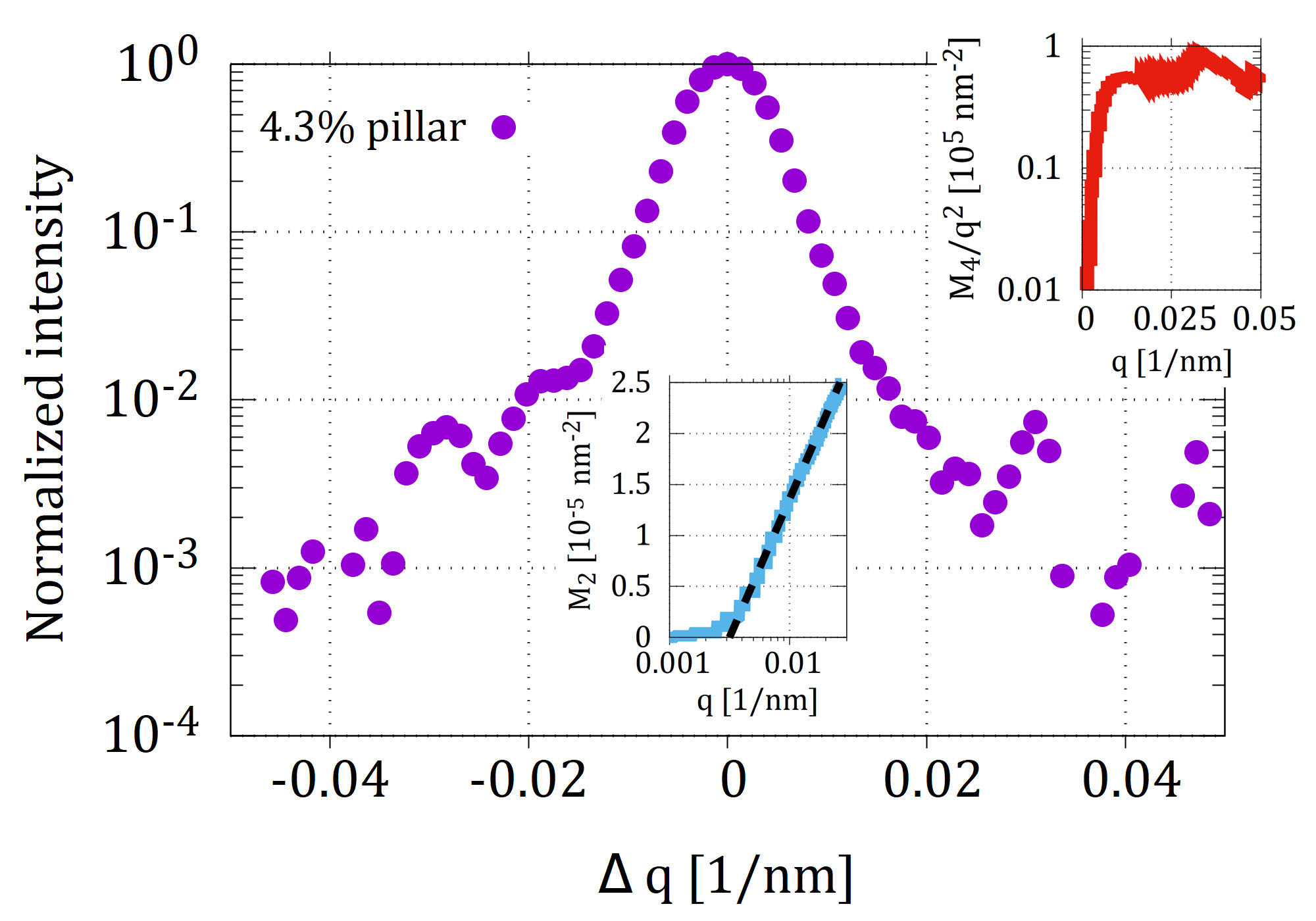} \\
\caption{\label{fig:05} XRD line profile measured on the 4.3\% deformed pillar with the calculated second ($M_2$) and fourth ($M_4$) order restricted moments. According to the theory \cite{Groma1998, Borbely2001}, if the broadening is caused by dislocations then $M_2$ must be logarithmic and $M_4/q^2$ must saturate as a function of the reciprocal space coordinate $q$ with prefactors proportional to the statistically stored dislocation density. The insets show that the experiments are in accordance with the expectations. (colours online)}
\end{center}
\end{figure}

When the 3D measurement was conducted on half of the 4.3\% pillar, the other half was lifted out from the original bulk sample and placed on top of a tungsten needle for X-ray line profile measurement. XRD experiment was carried out at the P21.2 beamline of PETRA III synchrotron in Hamburg, Germany. The sample was illuminated by a 67.4 keV monochromatic parallel beam and transmission diffraction images were recorded by a VAREX XRD4343CT area detector placed 3.025 m behind the specimen with the plane perpendicular to the direct beam. Pixel size of the detector was 150 $\mu$m, which results in an approximately $0.0028^{\circ}$ angular resolution in $2\theta$ diffraction angle, or 0.0027 1/nm resolution in the $q=2 \sin(\theta)/\lambda$ reciprocal space coordinate. $\lambda$ is the wavelength of the X-ray beam. Sample was rotated around its vertical axis $(\omega)$ until the Bragg condition was fulfilled for the $(-111)$ reflection and a diffraction spot appeared on the detector. Sample was then swinged $\pm 1.3^{\circ}$ in $\omega$ around the ideal reflection position and diffraction images were recorded. 4771 consecutive images were recorded with 1 s exposure time. Intensities were summed up for all images and also along the relevant section of the constant $2\theta$ lines (Debye-Scherrer ring) in order to achieve a good signal-to-noise ratio. The obtained radial line profile was then analysed by the method of restricted moments as described in Refs. \cite{Groma1998, Borbely2001}, and plotted in Figure \ref{fig:05}.

TEM measurements were conducted by a JEOL 200CX TEM at 200kV to verify the dislocation structure evolved as a result of external compression of the pillars.

\section{Results}

\subsection{$0.7\%$ deformation}
\begin{figure}[!ht]
\begin{center}
\includegraphics[width=0.4\textwidth]{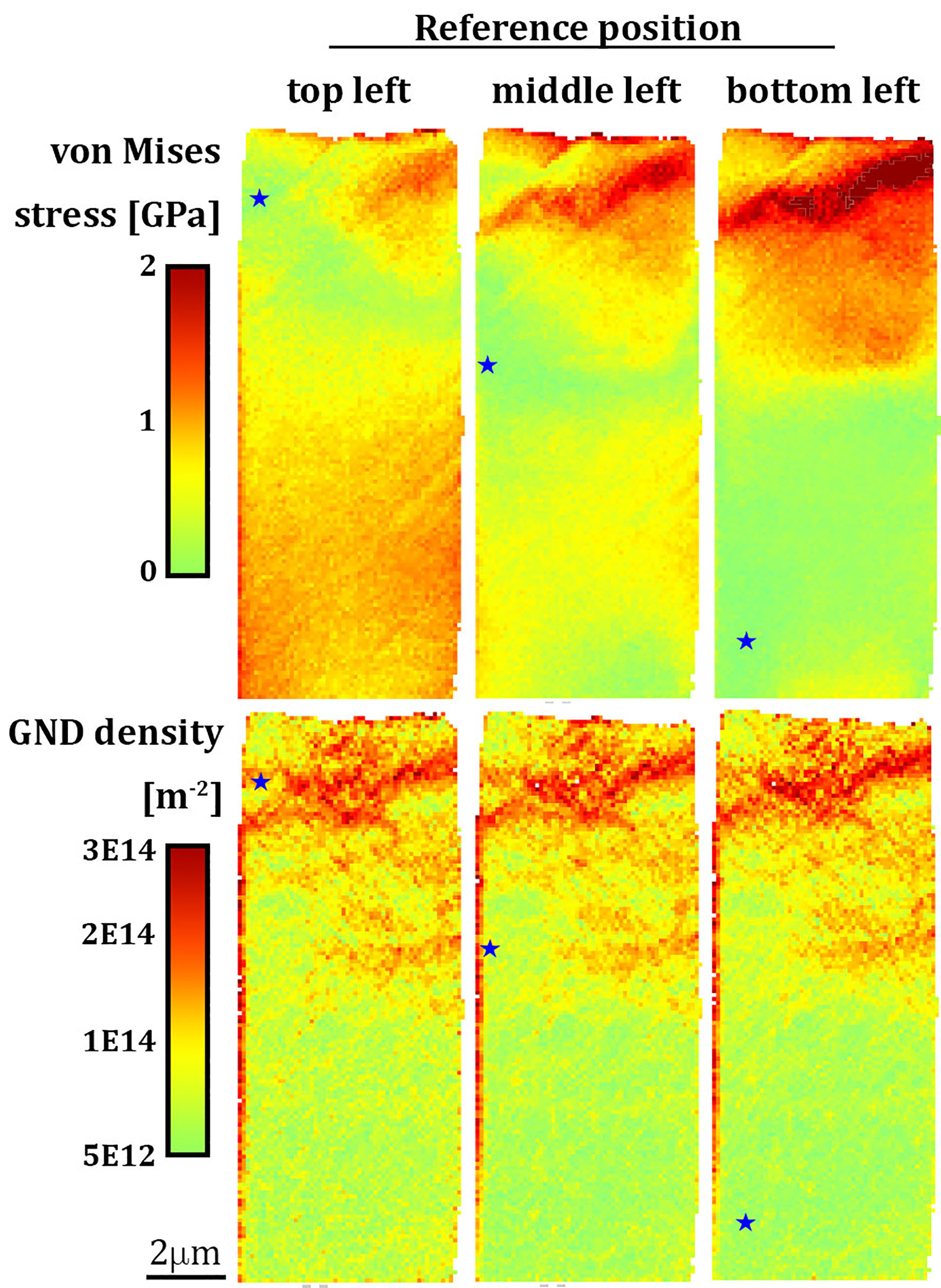} \\
\caption{\label{fig:06} The effect of different references shown in the 0.7\% deformation case. Three references were chosen from the top, middle and bottom region of the map (highlighted with a star sign). Two rows contain the calculated von Mises stress and GND density values for the same map. (colours online)}
\end{center}
\end{figure}

To demonstrate the effect of choosing different references from the same map, Figure \ref{fig:06} summarizes the von Mises stress and total GND density values calculated for three arbitrarily chosen reference patterns in case of the $0.7\%$ deformed pillar. The positions of the reference patterns are shown by stars on the maps. References were positioned at the left side of the pillars on the top, middle and bottom segment of the map. Reference patterns from the middle and right side were also used to conduct the same evaluation, and as they all show similar features, only the left side evaluation is presented here. The results show that, while the relative stress distribution in the pillar varies from one reference pattern to the other, the GND distribution is hardly affected by the reference pattern position. Choosing the reference pattern away from the strain gauge or using a simulated pattern \cite{alkorta2017} is necessary for absolute strain and stress analysis by HR-EBSD, but not crucial for the subject of this work, which concentrates on the distribution of GNDs.

\begin{figure}[!hb]
\begin{center}
\includegraphics[width=0.35\textwidth]{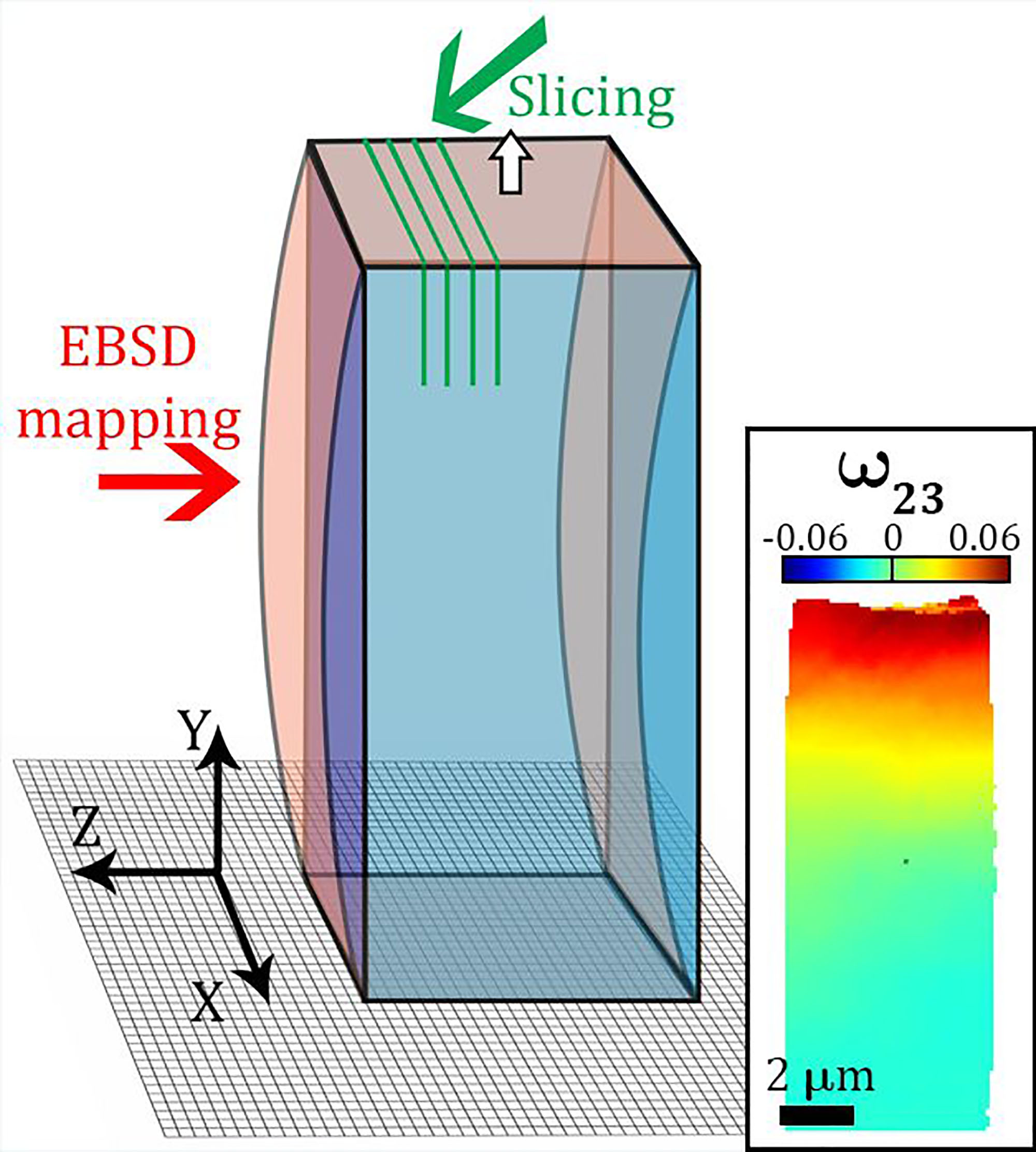} \\
\caption{\label{fig:07} Lattice distortion based on $\omega_{23}$, measured on the (0.7\%) deformed pillar. EBSD mapping direction (red arrow) and FIB slicing direction (green arrow) is indicated. The blue cuboid represents the undeformed micropillar, while red part shows the change in the shape of the deformed volume. The applied coordinate system and the normal of the compressed surface was also depicted. Inset shows the most significant rotation tensor component $\omega_{23}$ calculated for the first slice in radians. (colours online)}
\end{center}
\end{figure}

For the analysis of the slices measured on the same micropillar, a reference pattern from the middle of the first slice was selected. This reference pattern is then inserted to all measured slices at the same place to make the evaluation unitary. $N=38$ slices were measured on the 0.7\% deformed pillar, which means that a total thickness of 3.42 $\mu$m was mapped by HR-EBSD. The investigated volume was calculated by the difference in thickness of the pillar between the first and the last slice with ImageJ $(d_{\textrm{first slice}} - d_{\textrm{last slice}})$. The results are matching well with the calculated volume based on how many slices were made $(d_{\textrm{one slice}} \times N)$. This means that the FIB cut was precise, only the in-plane alignment of the HR-EBSD maps was necessary before creating the 3D model.

\begin{figure*}[!hb]
\begin{center}
\includegraphics[width=0.95\textwidth]{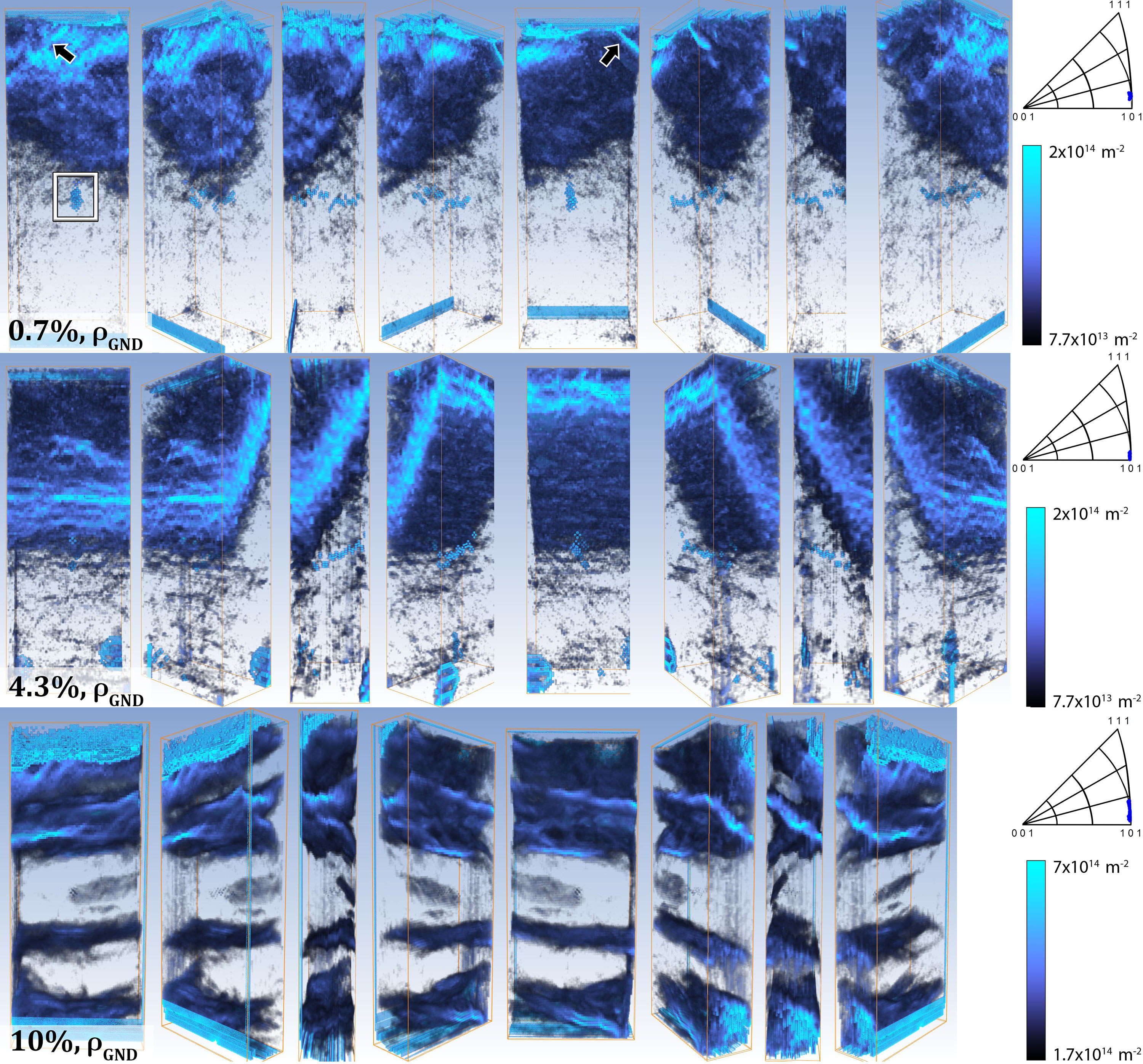} \\
\caption{\label{fig:08} 3D models of GND density values for the three micropillars (top row: $0.7\%$, middle row: $4.3\%$, bottom row: $10\%$) rotated around for inspection. First view in each row was made as we look directly on the outer surface of the pillar (first slice). The flat punch tip -- $Pt$ cap interface is located at the top of each model. Black arrows in the $0.7\%$ deformed pillar mark the FIB milling artefact. White square indicates the reference pattern replacement artefact. Absolute values of angles of the active slip systems can be easily identified. IPF figures on the right side show the orientation distribution viewed from the compression direction, plotted for one slice from the middle of each pillars. (colours online)}
\end{center}
\end{figure*}

To understand how the pillar changed after compression, elastic strain ($\varepsilon_{ij}$) and rotation ($\omega_{ij}$) components were calculated. The deformation was mostly localized on the top of the pillar. It should be noted that at the beginning of a compression test, if the top of the pillar and the surface of the  flat punch tip is not exactly aligned, the crystal lattice of the sample will rotate to have the same planar direction. This effect can be seen in the most significant rotation tensor component, $\omega_{23}$ in Figure \ref{fig:07}.

Pillar rotation corresponding to this direction was also interpreted. Other $\omega_{ij}$ components are not as conspicuous as the rotation around the $X$ axis. Figure \ref{fig:07} also shows FIB slicing and EBSD mapping directions. Highest values of $\omega_{23}$ correspond well to the initial lattice misalignment from the ideal (100) orientation mentioned in the previous section. 

The reconstructed 3D distribution of the GND density ($\rho_{GND}$) can be seen in Figure \ref{fig:08} top row. Inhomogeneous lattice rotation and GND accumulation close to the flat punch tip--pillar interface is observed. As the pillars used in the investigation are single crystalline, there are no limiting geometric constraints for dislocation motion in the system (if we can assume minimal surface damage from FIB milling). This means that the lattice can rotate freely and dislocations can slip out to the free surface, if they do not get tangled or pinned down by other dislocations. During compression at the first stage of deformation, GNDs are mainly formed close to the top of the pillar, along the active $\{111\}<110>$ type slip planes. Absolute values of angles of the active slip system ($35^{\circ}$ on the (100) face  and $55^{\circ}$ on the (011) face) can be identified on the 3D models. We observe only one activated slip system due to the aforementioned misalignment of the pillar. GNDs close to the Pt cap are generated to accommodate the lattice rotation at small strains. The misalignment is therefore responsible for the noticed inhomogeneous distribution of GND density.

The inverse pole figure (IPF) image in Figure \ref{fig:08} calculated for only one slice from the middle of the pillar reveals the beginning of the tendency for reorientation towards the (111) crystallographic orientation.

\subsection{$4.3\%$ deformation}

For the $4.3\%$ deformed pillar, $N=23$ slices were made, so that the total thickness of 3.10 $\mu$m was mapped by HR-EBSD. 3D map of the GND density can be seen in Figure \ref{fig:08} middle row. The GND density distribution clearly highlights the \{111\} type slip plane, but the direction differs from the primary slip system recognized in the $0.7\%$ deformed pillar. This slip system is also visible on the surface of the pillar by secondary electron imaging, and it was activated after the hardening (see in the supplementary video). At this deformation stage, multiple slip will be initiated after the preliminary lattice rotation  has been completed due to misalignment. Highest values of GND density are still localized at the top half of the micropillar, but their distribution is more spread than what we observed at the previous deformation step. In this case there is no observable FIB milling artefact. The deposited nanocrystalline Pt cap on the top allowed slicing without any major curtaining. 

The IPF orientation distribution is still localized at the (100) crystallographic direction, but it is slightly broader than at the previous deformation step.

The sample was then studied by X-ray line profile analysis. Values of $\Lambda \times \rho_{XRD}= 2.305 \times 10^{14}$ $1/m^2$ and $\Lambda  \times \rho_{XRD}=2.306 \times 10^{14}$ $1/m^2$ were obtained from the second and the fourth order restricted moments, respectively \cite{Groma1998}, where $\rho_{XRD}$ is the total dislocation density and $\Lambda$ is a geometrical factor depending on the dislocation type, diffraction vector, and the elastic constants. During compression test, the specimen was oriented for multiple slip, where four slip systems can be activated. Assuming that these slip systems are equally populated by straight edge and screw dislocations, the average $\Lambda$ contrast factor is 0.2552. Using this value we get $\rho_{XRD}=(9.03 \pm 0.9)\times 10^{14}$ $1/m^2$ dislocation density. Depending on which type of dislocations are activated, by adjusting the $\Lambda$ factor we can get a lower estimate for the total dislocation density of $4 \times 10^{14}$ $1/m^2$. This result supports our findings by HR-EBSD, that dislocations are present in the micropillar in elevated numbers, and they do not get eliminated on the free surfaces.

\subsection{$10\%$ deformation}

\begin{figure}[!ht]
\begin{center}
\includegraphics[width=0.35\textwidth]{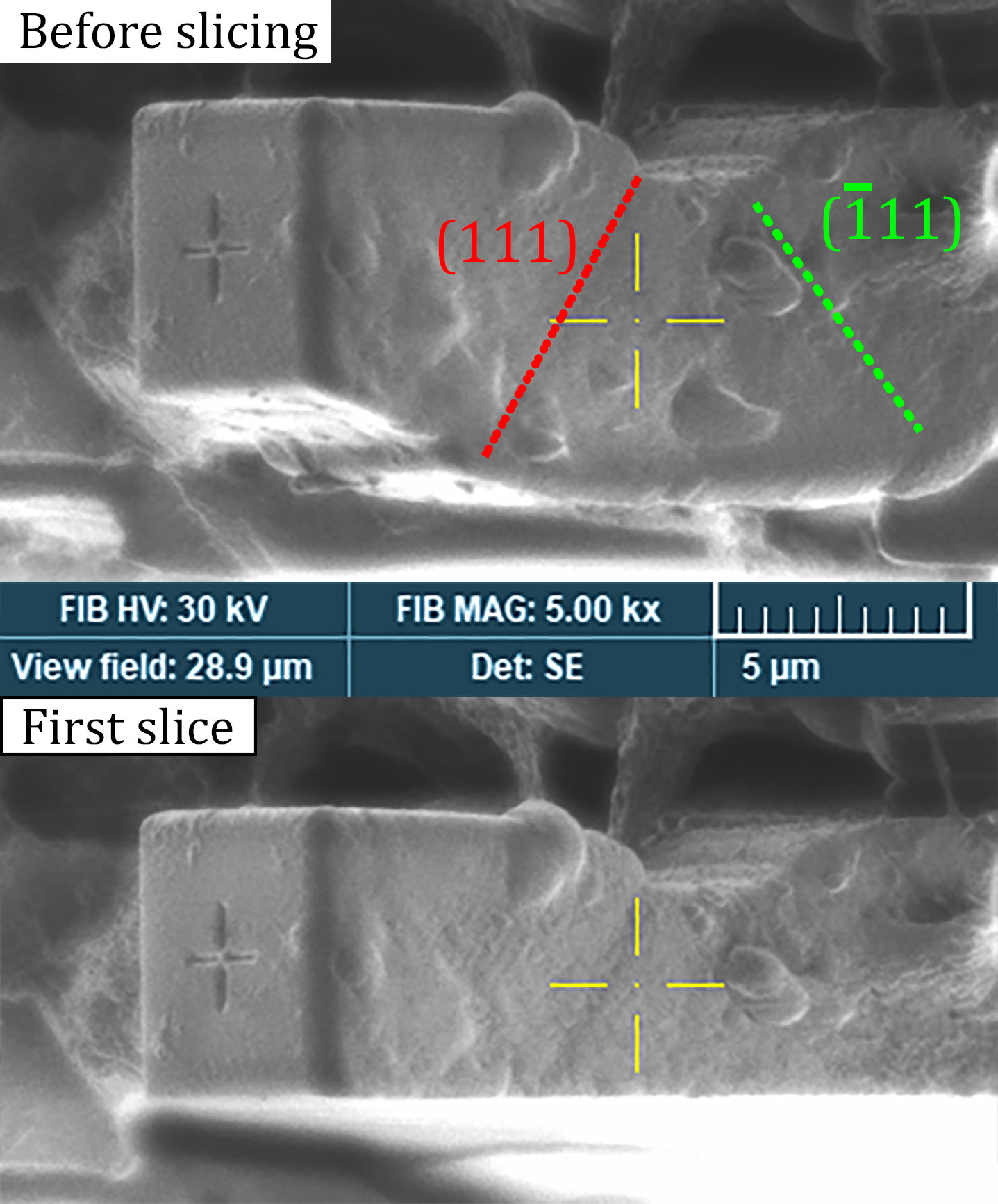} \\
\caption{\label{fig:09} FIB preparation of the first slice of the 10\% micropillar. External slip traces are also highlighted. Surface contamination occurred during the transportation of the sample which had no effect on the measurement whatsoever, as the surface of the pillar was repolished. Images were taken from the FIB view.}
\end{center}
\end{figure}

For the $10\%$ deformed pillar, $N=26$ slices were made, which means that the total thickness of 2.34 $\mu$m was mapped by HR-EBSD. After deformation, slip traces appeared on the surface of the pillar. Side view of the pillar before and after the first FIB cut can be seen in Figure \ref{fig:09}. Slip traces on the surface are clear indication that many of the dislocations had already reached the pillar surface. Although these external slip traces can be useful for identifying some activated slip directions, they might not coincide with the internal distribution of GNDs that accomodate lattice rotations.

To reveal residual GND distribution, the 3D model of the pillar is shown in Figure \ref{fig:08} bottom row. The colour scale differs from the first two pillars for better visualization. We observe the highest GND density among all tested pillars as expected. GNDs are piling up along various planes. The distribution is very different from the first two pillars, as more slip systems are visible at such elevated deformation. The evolution of orientation distribution towards the [111] direction is clear at this stage in the IPF figure.

\section{Discussion}

\begin{figure}[!ht]
\begin{center}
\includegraphics[width=0.45\textwidth]{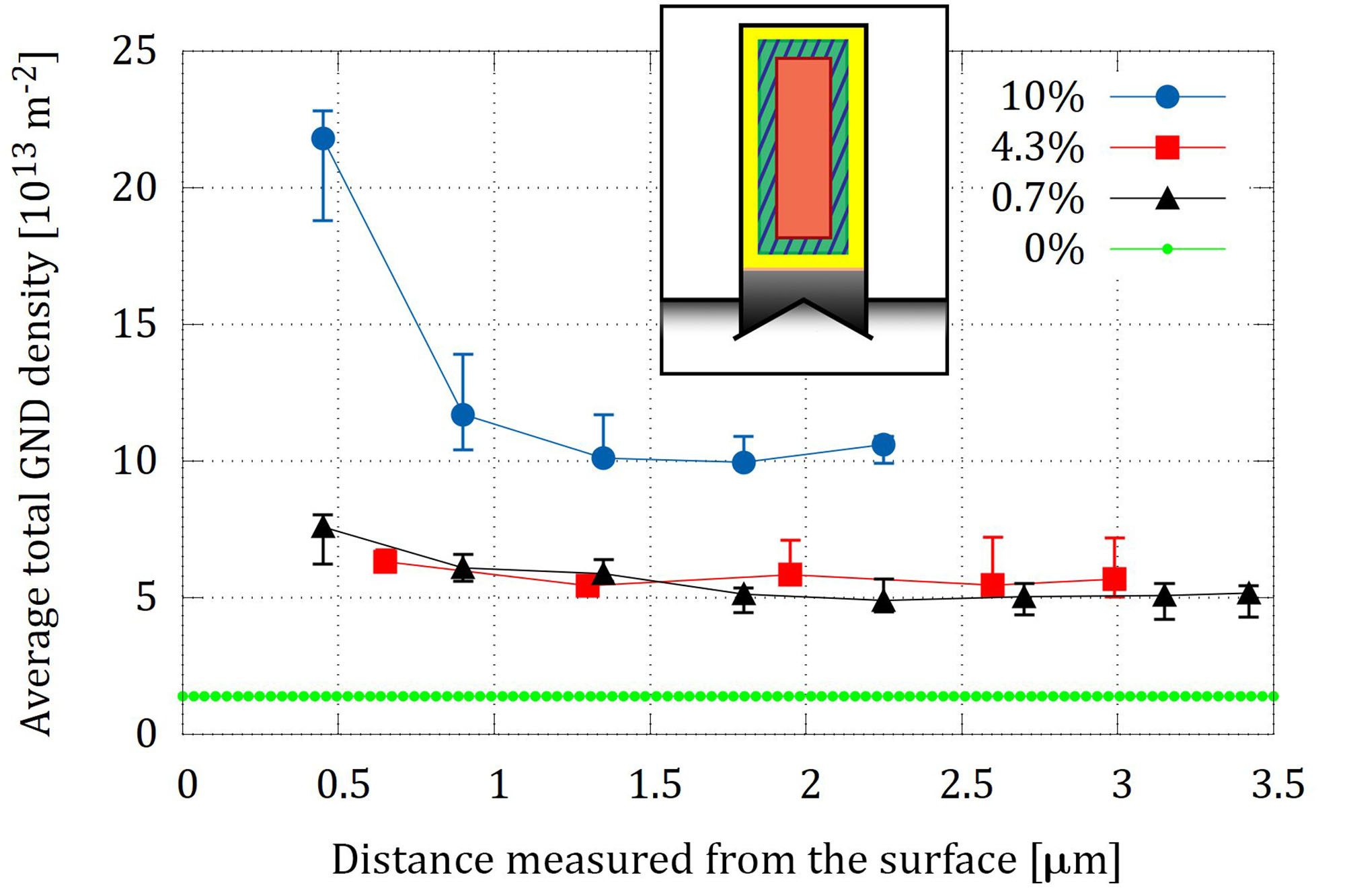} \\
\caption{\label{fig:10} Cross-sectional average GND density variation as a function of distance from the first slice on the surface. The inset shows schematics of the averaging areas selected for error estimation on each slice. Yellow area highlights the EBSD map taken from each slice. GND density values averaged over the orange area correspond to the lower estimate of error. The green squashed area and the orange area together show where average GND density values were calculated.  (colours online)}
\end{center}
\end{figure}

To study the average GND density evolution as a function of strain, slices were chosen from various cross-section location of the pillars. Figure \ref{fig:10} shows the total (edge + screw) GND density values averaged over a specified area calculated for all slices as a function of depth in the pillar ($=$ slice number $\times$ slice thickness). $\rho_{GND}$ values are calculated by CrossCourt software for all pixels within an EBSD map. Pixels close to the side of the pillar can have higher GND density values due to the decrease in pattern quality (thus increase in noise and uncertainty in cross-correlation calculation). The error from this uncertainty can be calculated by choosing different integration areas, as seen in the inset. The yellow area includes the total mapped surface. Green square generated from the yellow area by subtracting 10 pixels from each side represents the average $\rho_{GND}$ values. The yellow area gives the highest $\rho_{GND}$ values, while the orange area (generated from the green area by reducing each side by another 10 pixels) usually corresponds to smaller $\rho_{GND}$ values, as we tend to miss dislocations piling up close to the top of the pillar. The values plotted in Figure \ref{fig:08} and \ref{fig:10} were calculated by the aforementioned $L^1$ optimisation scheme.

Baseline ("0\%") was established on a FIB prepared surface close to the pillars. A small area was polished by a beam with FIB settings identical to the pillar slicing parameters. The resulting $1.38 \times 10^{13}$ $1/m^2$ value is typical for the limitation of the cross-correlation based GND density determination method. After deformation, average $\rho_{GND}$ values were higher but fairly constant throughout the depth of each pillar. $0.7\%$ and $4.3\%$ pillars have similar magnitudes of $\rho_{GND}$ ($\approx 5.7 \times 10^{14}$ $1/m^2$). In case of the $10\%$ sample, average $\rho_{GND}$ values are higher close to the surface than in the middle of the pillar. This can be interpreted with the help of the 3D map in Figure \ref{fig:08} bottom row. High values of $\rho_{GND}$ are located close to the interface between the flat punch tip and the $Pt$ cap. Due to the geometrical restriction and misalignment of the pillar, coupled with the hard $Pt$ cap that was deposited on the top, GNDs that are continuously generated cannot exit the system through the top surface. This small area will contain the highest values of stresses. As we explore the inner regions of the pillar, the high $\rho_{GND}$ volume is reduced and we receive similar levels of average GND density with less error throughout the rest of the thickness ($\rho_{GND}^{10\%} \approx 1.2 \times 10^{14}$ $1/m^2$). To better understand dislocation density in such small volumes, $\rho$ of $10^{14}$ $1/m^2$ value corresponds to an average distance of 100 nm between dislocations. This is $\sim 60$ times smaller than the size of the pillar. 

Dislocation density values can be compared with an earlier study by Kal\'acska \emph{et al.} \cite{Kalacska2017} where bulk single crystalline samples were deformed by channel-die compression. The calculated total dislocation densities were measured on the surface of the samples by X-ray diffraction along with $\rho_{GND}$ values by HR-EBSD. Dislocation densities from both studies are summarized in Table \ref{table:02}. 

\begin{table}[!ht] 
\begin{tabular}[width=\textwidth]{|l|l|l|}\hline 
Sample & $\rho_{XRD}$ $(1/m^2)$ & $\rho_{GND}$ $(1/m^2)$\\
\hline\hline
Bulk $6\%$ \cite{Kalacska2017} & $7.3 \times 10^{14}$ & $2.3 \times 10^{14}$\\ 
Bulk $10\%$ \cite{Kalacska2017} & $1.5 \times 10^{15}$ & $1.3 \times 10^{15}$\\ \hline
pillar $0.7\%$ & -- & $5.6 \times 10^{13}$\\ 
pillar $4.3\%$ & $(4-9) \times 10^{14}$ & $5.7 \times 10^{13}$\\ 
pillar $10\%$ & -- & $1.2 \times 10^{14}$\\ 
\hline

\end{tabular} 
\centering \caption{Dislocation density comparison measured on bulk copper single crystal samples and micropillars.} 
\label{table:02} 
\end{table}

\begin{figure}[!ht]
\begin{center}
\includegraphics[width=0.45\textwidth]{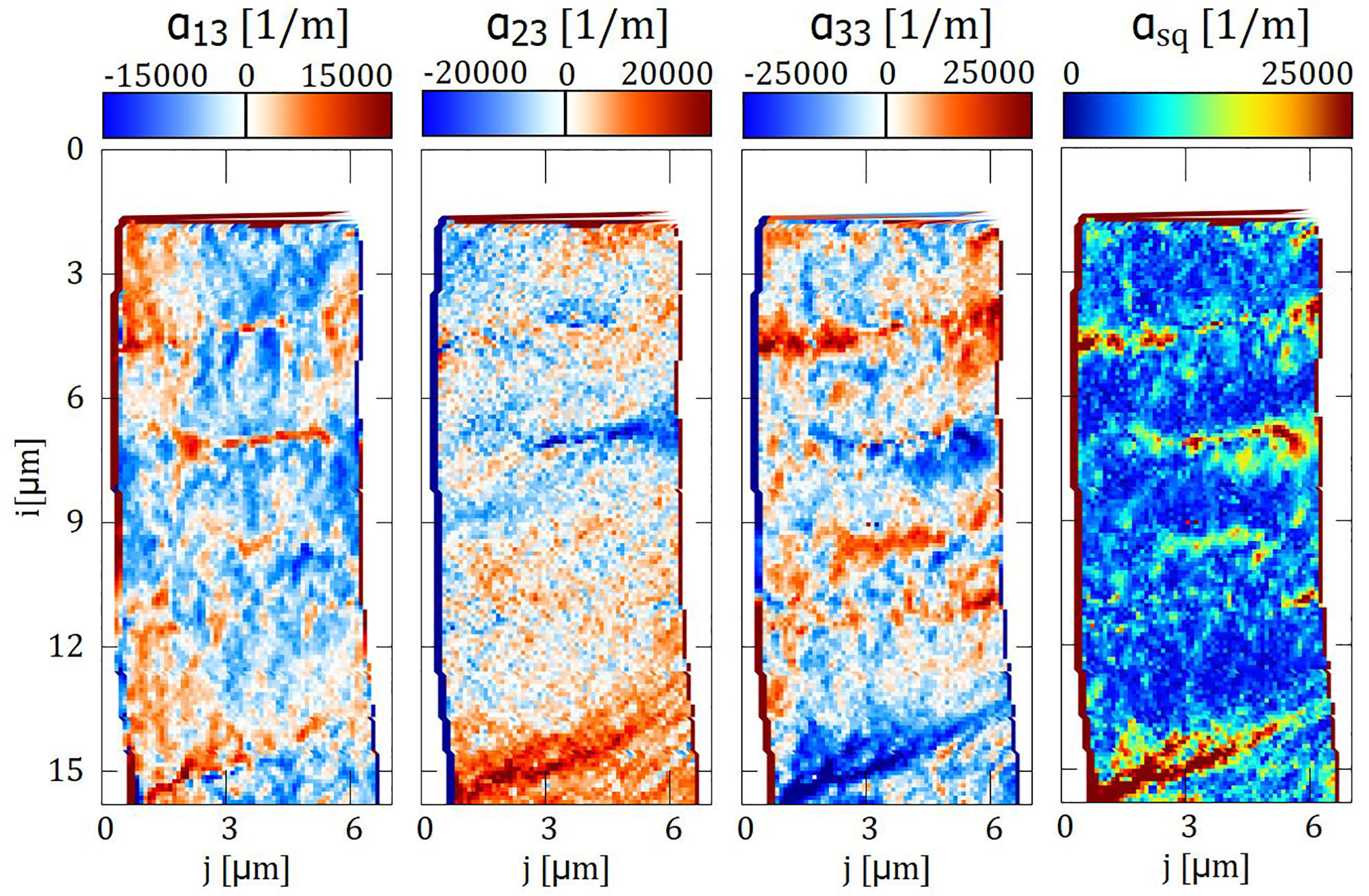} \\
\caption{\label{fig:11} $\alpha_{i3}$ and $\alpha_{sq}$ values plotted for the last (26th) slice of the 10\% deformed pillar. (colours online)}
\end{center}
\end{figure}

\begin{figure*}[!ht]
\begin{center}
\includegraphics[width=0.9\textwidth]{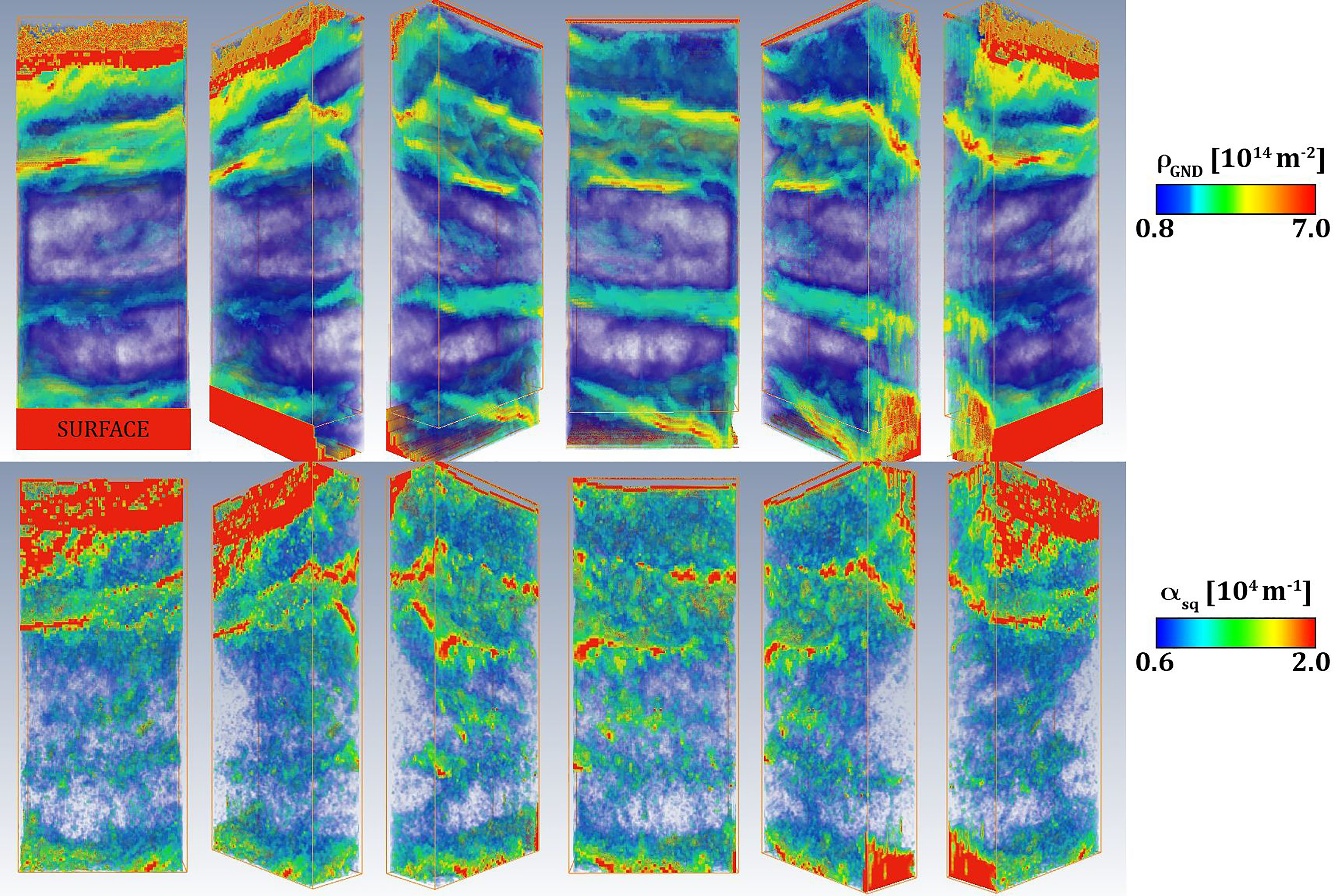} \\
\caption{\label{fig:12} 3D GND density distribution calculated by applying the $L^1$ optimisation vs. distribution of $\alpha_{sq}$ measured on the 10\% pillar (colours online).}
\end{center}
\end{figure*}

\begin{figure*}[!ht]
\begin{center}
\includegraphics[width=0.95\textwidth]{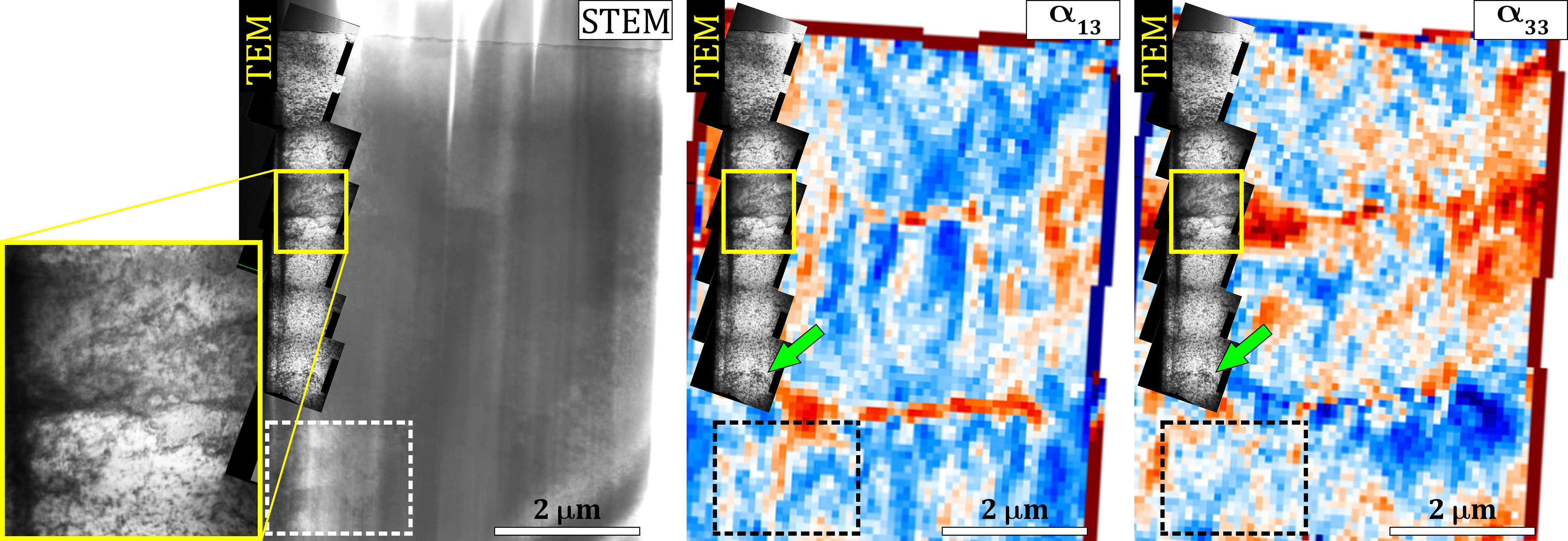} \\
\caption{\label{fig:13} STEM, TEM and HR-EBSD results scaled together to link actual dislocation distribution to features on the $\alpha_{i3}$ maps. The inset shows a magnified area on the TEM map. Dotted line circles the area where dislocation cell structure is visible on both STEM and $\alpha_{i3}$ maps. Green arrow points out similar features on both TEM and $\alpha_{i3}$ maps. $\alpha_{i3}$ colour scales are identical to Figure \ref{fig:11}. (colours online)}
\end{center}
\end{figure*}

$\rho_{XRD}$ values calculated for these micropillars correspond well to earlier results.  On the other hand, $\rho_{GND}$ values are one order of magnitude less in micropillars than previously measured by HR-EBSD on bulk samples. This can be explained by the difference in the deformed volume and the presence of free surface, letting dislocations to escape the system during deformation. At such small sample volumes the surface effect cannot be neglected. The behaviour of the material resembles to the bulk but it has very distinct properties. Also, during uniaxial micropillar compression only a small amount of the generated dislocations is geometrically necessary. 

Another method to study the distribution of GNDs is based on the calculation of the individual GND density tensor components. $\alpha_{i3}$ were calculated from CrossCourt's $\varepsilon_{ij}$ and $\omega_{ij}$ by a C++ code. $\alpha_{i3}$ and $\alpha_{sq}=\sqrt{\alpha_{13}^2 + \alpha_{23}^2 + \alpha_{33}^2}$ values are plotted in Figure \ref{fig:11}. $\alpha_{sq}$ is proportional to the GND density (Equation \ref{eq:gnd_ai3}).

$\alpha_{sq}$ values were calculated for all slices measured on the 10\% pillar. 3D maps highlighting the distribution of GNDs determined by the two methods are then compared in Figure \ref{fig:12}. Main features on both models appear to be similar, although fine details seem to be missing on the $\rho_{GND}$ reconstruction. Because of the additional assumptions that are made during $L^1$ optimisation, the cell structure of GNDs only appear on the $\alpha_{sq}$ model. The application of the $L^1$ optimisation is therefore useful for the estimation of GND density value for the whole investigated sample, but it can lead to blurred distributions. $\alpha_{sq}$ values on the other hand are exact, therefore they are more suitable to study the true distribution of GNDs.

The advantage of determining the $\alpha_{i3}$ components is that they are signed values, therefore we can distinguish sub-structures based on the sign of $\alpha_{i3}$. Furthermore, the determination of $\alpha_{i3}$ values are unequivocal, no application of a further optimization method is necessary that would alter the distribution of the GND density. 
In order to relate the actual dislocation distribution to $\rho_{GND}$, the remaining volume of the 10\% deformed pillar was lifted out and a TEM lamella was prepared by FIB. The distance between the lamella and the last HR-EBSD slice was about (100-200) nm, so that the features in all three maps measured by different techniques can be related. After the FIB preparation of the lamella, scanning transmission electron microscopy (STEM) technique was used to check the distribution of dislocations on the whole sample. Images measured with all three techniques (STEM, TEM and HR-EBSD) have been summarized in Figure \ref{fig:13}.

FIB milling artefacts can be easily seen in the STEM image (vertical lines). STEM has lower resolution than conventional TEM but it has bigger field of view. Dotted line circles the area where dislocation cell structure is visible on both STEM and $\alpha_{i3}$ maps. An inset taken from the TEM map and a green arrow points out the dislocation pile-ups where elevated values of $\alpha_{13}$ and $\alpha_{33}$ maps appear. TEM and STEM images confirm the formation of dislocation cells inside the micropillar. Features on all three maps correspond well to each other.

It is shown that 3D HR-EBSD is capable of determining GND distributions in compressed micropillars. These results can be used as input for comparing the evolution of dislocation microstructure obtained experimentally and by dislocation dynamics and crystal plasticity modelling. Micropillar compression requires new physics, therefore any new information on dislocation mechanisms will add a piece to the puzzle of understanding non-deterministic behavior and small-scale plasticity. Our article starts with the simplest possible system, as face-centered cubic single crystal copper has been widely studied. This work is just the beginning of a series of following studies that will focus on different geometries and materials.

\section{Summary and conclusion}

In this paper we have introduced a new technique of 3-dimensional high resolution electron backscatter diffraction (3D HR-EBSD). Identical micropillars ($6$ $\mu m \times 6$ $\mu m \times 18$ $\mu$m in size) were created by FIB from annealed copper single crystal sample. Compression tests were stopped at different levels to investigate GND density distributions in 3D. HR-EBSD coupled with serial slicing successfully revealed the evolution of GND distribution. HR-EBSD reference pattern evaluation showed that strain distribution depends strongly on the choice of the reference, however GND density values were less affected by this issue. We concluded that pillar tops had a small misalignment compared to the flat punch tip prior to deformation. This resulted inhomogeneous distribution of GNDs in the micropillars. Dislocations piled up close to the flat punch tip at the early stage of deformation. Activated slip planes were identified, and average GND densities and total dislocation densities were calculated throughout the samples' cross-sections. The present dislocation cell structure was revealed by TEM imaging and also by HR-EBSD mapping. Comparison between earlier results measured on bulk samples show similar order of magnitude ($10^{14}$ $m^{-2}$) in total dislocation density. GND density evaluation from HR-EBSD on the other hand detected one order of magnitude less GNDs present in the system ($10^{13}$ $m^{-2}$). This is a consequence of the small volumes of micropillars and the presence of free surfaces where dislocation can exit the system by sliding out, leaving slip traces behind on the outside of the pillars. In the case of micropillars larger than 5 $\mu$m we confirmed that although some dislocations have left the system, but still enough remain to form cells as in bulk specimen, leading to work hardening. In accordance with earlier reports \cite{Zhao2019} this Taylor type work hardening drives a large part of the deformation mechanism in pillars larger than 1 $\mu$m.

GND density distribution calculated by the $L^1$ optimisation method was compared to the distribution of $\alpha_{i3}$ dislocation density tensor elements. Dislocation cell structure was observed on both $\alpha_{i3}$ and $\alpha_{sq}$ maps, providing the option to investigate deformation-influenced materials’ properties by 3D HR-EBSD in crystalline materials.

Overall, we found that at this intermittent scale the material can be considered neither bulk nor nano, we successfully connected our findings to earlier studies (for example GND density build-up in micropillars \cite{Kiener2011}, GND and total dislocation density comparison with bulk \cite{Kalacska2017}, and detected dislocation cell formation \cite{Zhao2019}). On the one hand, as a typical bulk phenomenon a complex GND structure evolves which leads to significant strain hardening during compression and the accumulation of dislocations in the system. The deformation is, therefore, governed by the collective dynamics of dislocations rather than the dynamics of individual dislocations (such as source-surface interactions). On the other hand, a significant number of dislocations can still leave the system through the pillar surface and, thus, the measured GND density values are an order of magnitude lower than typical bulk values. Consequently, a strong size effect can be observed: according to Fig.~\ref{fig:03} the yield stress of $\sim 150$ MPa is significantly higher than the bulk value. In addition, the observed large strain bursts are also characteristic to nano-scale, dislocation starved plasticity. So, the proposed 3D HR-EBSD method enabled us to experimentally study the role of GNDs during micromechanical testing, and opens new perspectives to comprehend small-scale plasticity of crystalline materials in more detail.

\section*{Acknowledgement}

The Authors greatly acknowledge the assistance of D\'{a}vid Ugi and Anik\'{o} N\'{e}met in participating in the micromechanical tests, \'{A}bel Szab\'{o} for assisting during micropillar FIB preparation, Alajos \"{O}. Kov\'acs for TEM imaging. This work was completed in the ELTE Institutional Excellence Program (1783-3/2018/FEKUTSRAT) supported by the Hungarian Ministry of Human Capacities. GI and PDI are grateful for financial support of the National Research, Development and Innovation Found of Hungary (project Nos. NKFIH-K-119561 and NKFIH-KH-125380). PDI is also supported by the \'UNKP-18-4 New National Excellence Program of the Hungarian Ministry of Human Capacities and by the János Bolyai Scholarship of the Hungarian Academy of Sciences. SzK was supported by the Hungarian Scholarship \'UNKP-16-3 New National Excellence Program of the Ministry of Human Capacities, No. ELTE/8495/53(2016) and by the EMPAPOSTDOCS-II programme, that has received funding from the European Union’s Horizon 2020 research and innovation programme under the Marie Skłodowska-Curie grant agreement number 754364. We acknowledge DESY (Hamburg, Germany), a member of the Helmholz Association HGF, for the provision of experimental facilities. Parts of this research were carried out at PETRA III facility, and we would like to thank Zolt\'{a}n Heged\H{u}s and Ulrich Lienert for assistance in using P21.2 beamline.

\section*{Data availability}

The raw/processed data required to reproduce these findings cannot be shared at this time as the data also forms part of an ongoing study. When the ongoing study is concluded, all raw/processed data will be uploaded to Mendeley Data repository.

\bibliography{mybibfile}

\end{document}